\documentclass[Journal,10pt]{IEEEtran}
\usepackage{cite}
\usepackage{threeparttable}
\usepackage{color}
\usepackage{float}
\usepackage{tabularx,colortbl}
\usepackage{setspace}
\usepackage{graphicx,amsmath,psfrag,bm}
\usepackage{subfigure}
\usepackage[usenames,dvipsnames]{xcolor}
\usepackage{mathabx}
\usepackage{cancel}
\usepackage{soul,xcolor}
\usepackage{amsmath}
\usepackage{cancel}
\usepackage{fixmath}
\usepackage{caption}
\usepackage[normalem]{ulem}
\usepackage{nicematrix}

\usepackage{tabularx}
\usepackage[symbol]{footmisc}
\usepackage{threeparttable}

\graphicspath{{Figs/}}

\ifCLASSINFOpdf
\else
\fi

\begin{document}
\setstcolor{red}
\title{Metasurface-Based Time-Reversal Focusing \\ for Brain Tumor Microwave Hyperthermia }
\author{Mohamad J. Hajiahmadi, \IEEEmembership{Student Member, IEEE}, Reza Faraji-Dana, \IEEEmembership{Senior Member, IEEE}, \\ and Christophe Caloz, \IEEEmembership{Fellow, IEEE}
\thanks{M. J. Hajiahmadi and R. Faraji-Dana are with The School of Electrical and Computer Engineering, College of Engineering, University of Tehran, 14395-515, Tehran, Iran (emails: mjhajiahmadi@ut.ac.ir, reza@ut.ac.ir).  C. Caloz is with KU Leuven, Kasteelpark Arenberg 10-box 2444, 3001 Leuven, Belgium (email: christophe.caloz@kuleuven.be).\\ }}

\maketitle

\begin{abstract}
We introduce a metasurface-based system for the microwave hyperthermia treatment of brain tumors, where the metasurface is designed by the time-reversal technique for optimal focusing at the location of the tumor. Compared to previous time-reversal microwave hyperthermia systems, which are based on antenna arrays, the proposed system offers two advantages: 1)~it provides superior focal resolution, due to the finer sampling capability of its deeply-subwavelength metasurface particles compared to that of typical array antenna elements, 2)~it features much lower complexity and cost, due to its direct illumination principle compared to the heavy feeding network of the array system. The system leverages reflecting metallic walls and a rich scattering structure in combination with the metasurface to achieve high hyperthermia performance with maximal simplicity. The metasurface system is demonstrated by full-wave simulation results, which show that it represents a viable novel noninvasive hyperthermia technology.
\end{abstract}


\section{Introduction}
\label{sec:introduction}

\IEEEPARstart{H}{yperthermia} therapy, or hyperthermia for short, is a medical treatment in which cancerous tumors are exposed to heat for sensitization to radiotherapy or chemotherapy~\cite{intro-hyperthermia1,intro-hyperthermia2,intro-hyperthermia3,intro-hyperthermia4,intro-hyperthermia5}. Specifically, microwave hyperthermia is a noninvasive type of hyperthermia where the heat is applied to the tumor across the body via electromagnetic radiation. This technology  is considered a promising competitor to ultrasound hyperthermia for brain tumors, given its superior penetration across the skull and potentially higher resolution.

Previously reported hyperthermia systems are based on antenna array technology~\cite{Microwave_Hyperthermia1, Microwave_Hyperthermia2, Microwave_Hyperthermia3, Microwave_Hyperthermia4, Microwave_Hyperthermia5}. This technology, which initially resorted to simple beam forming that could not well handle the complexity of inhomogeneous and irregularly shaped organs and tissue, strongly benefited from the time-reversal design technique~\cite{Time-reversal_for_HT1, Time-reversal_for_HT2, Time-reversal_for_HT4}, pioneered by Fink and colleagues~\cite{TR-Ultrasonic,TR-Microwave1,TR-definition,TR-Microwave,TR-definition1}. However, it still suffers from drawbacks inherent to the basis antenna array. First, the relatively large size and spacing ($\sim\lambda/2$) of the antenna elements several restricts the sampling finesse of the time-reversed field, which translates into a relatively small focal resolution and undesired hot spots in healthy regions~\cite{Time-reversal_for_HT3,Antenna_Complexity3} for a given system size. Second, the large number of required antenna elements ($\sim 12-64$) with simultaneous phase and amplitude control necessitates a complex and bulky feeding network~\cite{Antenna_Complexity1,Antenna_Complexity2, complex_feeding_network1, complex_feeding_network2}, which implies in turn a high cost.

In order to mitigate these issues, we propose here a metasurface-based microwave hyperthermia system, leveraging the unique wave front transformation capabilities of  metasurfaces~\cite{GSTC,general_MS,WFS_MS_scirep,SpatialProcessor_MS,THz_WFS_MS,WFS_MS_nature}, for the treatment of brain tumors. In order to make this system maximally simple and economical, we limit our design to a single flat metasurface assisted by reflecting metallic walls~\cite{TR-Cavity1,TR-Cavity2,TR-Cavity3,TR-Cavity4} and a scattering-enhancement structure~\cite{Scatterer-TR-focusing}.

The sequel of the paper is organized as follows. Section~\ref{sec:met_foc_syst} describes the proposed  metasurface system. Section~\ref{sec:TR_Foc_Des} outlines the time-reversal design procedure, while Sec.~\ref{sec:ms_susc_synth} derives the susceptibility formulas of the metasurface producing the reversed fields obtained by this procedure. Section~\ref{sec:hype_model} discusses the modeling aspects of the system. Section~\ref{sec:sys-design} details its overall design procedure and Sec.~\ref{sec:results} presents the related full-wave demonstration. Finally, Sec.~\ref{sec:conclusion} concludes the paper with a brief summary and discussion on possible improvements. \\

\section{Metasurface Focusing System}\label{sec:met_foc_syst}

\begin{figure} [h!]
\centering
\includegraphics[width=\columnwidth]{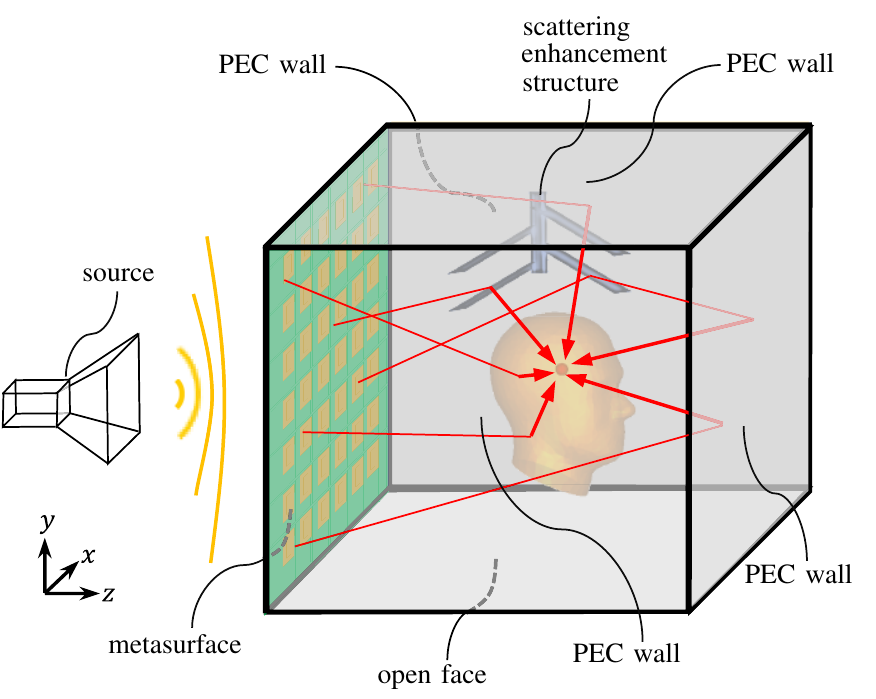}
		\caption{Proposed metasurface-based time-reversal focusing  system for the hyperthermia treatment of deep-seated brain tumors.}
\label{fig1_metasurface system}
\end{figure}

Figure~\ref{fig1_metasurface system} depicts the proposed metasurface-based time-reversal focusing system for the hyperthermia treatment of deep-seated brain tumors. The system consists of the following parts: 1)~an antenna source, 2)~the metasurface, 3)~a cubic cavity with the face on the side of the source formed by the metasurface, the other three lateral faces and the top face formed by metal plates, and the bottom face open to pass the head of the patient, 4)~a scattering enhancement structure.

The system works as follows. The antenna source, assumed to be placed in the far-field of the cavity and directed perpendicular to it, uniformly illuminates the metasurface. The metasurface, matched to the wave incoming from the antenna, scatters this wave into the cavity, loaded by the patient's head, in a way that is optimized for focusing the energy of the wave at the point of the tumor, according to the time-reversal design principle that will be described in Sec.~\ref{sec:TR_Foc_Des}. The focusing process is assisted by the metal walls, assumed to operate as Perfect Electric Conductors (PECs), and by the scattering enhancement structure, whose exact shape and operation will be discussed later.

\section{Time-Reversal Design Principle}\label{sec:TR_Foc_Des}

As mentioned in Sec.~\ref{sec:met_foc_syst}, the metasurface must be designed in such a manner that it optimally focuses the energy provided by the source to the tumor position in the head of the patient. Given the complexity of the shape of the head and of its bone and tissue media, such a design seems, a priori, to represent a formidable task. Fortunately, the concept of time-reversal focusing~\cite{TR-definition1,TR-Microwave1,TR-Ultrasonic,TR-definition} provides a solution to meet this challenge.

\begin{figure} [h!]

\centering

\includegraphics[width=\columnwidth]{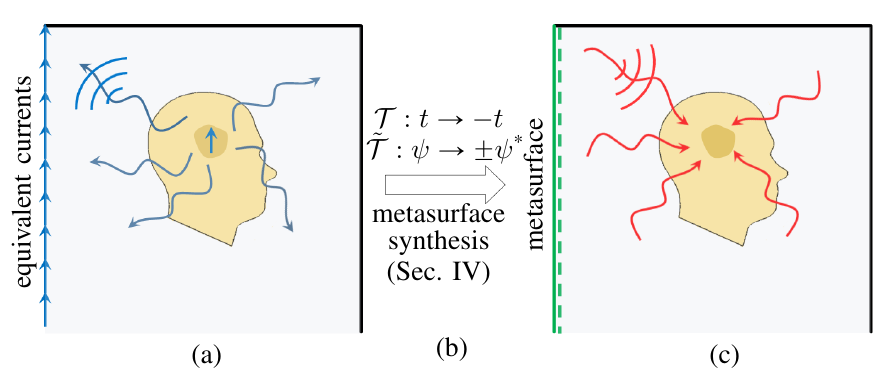}	
		\caption{Time-reversal focusing design for the metasurface system in Fig.~\ref{fig1_metasurface system}, consisting in the following three successive steps: (a)~Forward-propagation simulation. (b)~Mathematical time reversal ($\mathcal{T}$) or electromagnetic phase conjugation ($\tilde{\mathcal{T}}$). (c)~Backward propagation activation of the system.}

\label{fig2_TR-principle}

	\end{figure}

Figure~\ref{fig2_TR-principle} describes the time-reversal design procedure that we shall follow here, using the three usual successive steps inherent to time-reversal-based systems:

\begin{enumerate}
    \item See Fig.~\ref{fig2_TR-principle}(a) -- ``Forward Propagation'' simulation: assuming an appropriate electromagnetic model of the human head and using a full-wave simulator,
    i)~place a small dipole\footnote{\label{fn:multipole_mod}We assume in this paper that the tumor is small enough with respect to the wavelength and simple enough in terms of its composition so that it can be modeled by a simple \emph{dipolar polarization current}~\cite{collin1990field}. Larger and/or more complex tumors, or other types of focusing structures in other applications, may require more extensive modeling using volume polarization currents~\cite{PolarizationCurrents_Sarkar}.} at the desired focal point corresponding to the position of the tumor, assumed to be known from prior Magnetic Resonance Imaging (MRI); ii)~compute the electromagnetic field radiated by this dipole through the head and the surrounding cavity; iii)~record the corresponding scattered fields on the metasurface side within the cavity, and compute the equivalent, or Huygens currents;
    \item See Fig.~\ref{fig2_TR-principle}(b) -- Reversal and Synthesis: i)~time-reverse or, equivalently in the frequency domain, electromagnetically phase-conjugate\footnote{\label{fn:tr}Time reversal (in the time domain) is not exactly equivalent to phase conjugation in the frequency domain. Phase conjugation in the frequency domain is exactly equivalent to time reversal in the time domain insofar as the phase of waves is concerned. However, time reversal also implies the reversal of the sign of the magnetic field, due its odd time-reversal nature, without change of the sign of the (even) electric field~\cite{Ishimaru_1990}. We consequently refer here to \emph{electromagnetic phase conjugation} as phase conjugation plus the change of the sign of the magnetic field.} these currents. ii)~synthesize the metasurface so that it transforms the wave received from the antenna source in Fig.~\ref{fig1_metasurface system} into the required reversed/conjugated Huygens current distribution at its output (within the cavity) using the GSTC technique that will be described in Sec.~\ref{sec:ms_susc_synth};
    \item See Fig.~\ref{fig2_TR-principle}(c) -- ``Backward Propagation'' activation: activate the system (Fig.~\ref{fig1_metasurface system}), i.e., switch the antenna source on, which will produce the desired energy focusing at the position of the tumor according to the time reversal principle.
\end{enumerate}

According to the time reversal principle~\cite{TR-definition1,TR-Microwave1,TR-Ultrasonic,TR-definition}, the field radiated by the dipole in the forward-propagation simulation step should be perfectly replicated in the backward propagation activation step after proper time reversal, except for a decrease in the field intensity due to the time-reversal asymmetric nature of loss~\cite{Caloz_PRAp_10_2018}, which can be simply compensated by tuning the source power in the present application. However, such ``perfect'' focusing requires, according to the Huygens principle~\cite{Ishimaru_1990,general_Huygens}, that the reversed Huygens currents be defined on a \emph{complete surface} around the region of interest~\cite{TR-Ultrasonic}. This condition is certainly not satisfied in the system of Fig.~\ref{fig1_metasurface system}, where only one face of the cubic cavity reverses the Huygens currents, in order to minimize the system's complexity\footnote{One may for instance conceive, at the cost of dramatically increased complexity, a system with four additional time-reversing metasurfaces, forming the other three lateral walls and the top wall of the cavity, and illuminated by corresponding antenna sources, or a spherical kind of ``helmet'' with judiciously placed external sources.}, and where the bottom face must be left open for obvious reasons. Fortunately, the penalty for violating the complete reversal surface condition can be mitigated by enriching the scattering diversity of the cavity with reflecting walls and scattering elements~\cite{TR-Cavity1,TR-Cavity3}, which is the reason behind the presence of these elements in Fig.~\ref{fig1_metasurface system}.

\section{Metasurface Susceptibility Synthesis}\label{sec:ms_susc_synth}

The problem at hand at this point is thus to design the metasurface in Fig.~\ref{fig1_metasurface system}, i.e., a metasurface transforming the field incident from the antenna source into a transmitted field that is identical to the field that would be radiated by the time-reversed Huygens currents obtained from the procedure of Sec.~\ref{sec:TR_Foc_Des}.

\begin{figure} [h!]
\centering

\includegraphics[width=0.7\columnwidth]{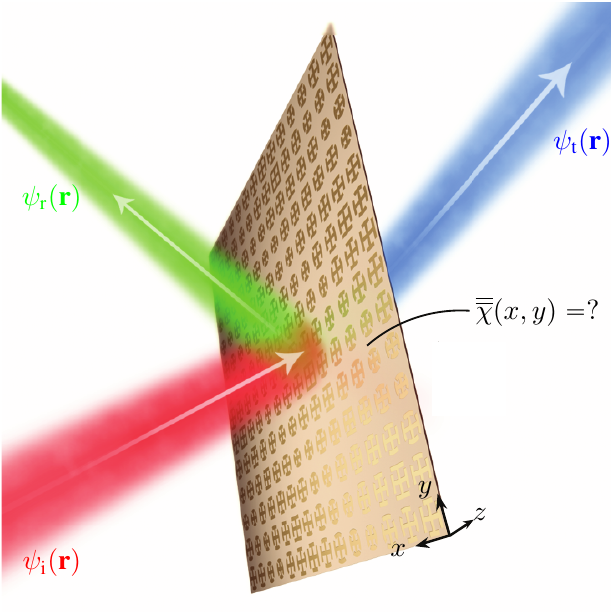}
\caption{General metasurface synthesis problem: determine the surface susceptibility function $\overline{\overline{\chi}}(x,y)$ that transforms the specified incident field, $\psi_\text{i}(\mathbf{r})$ into the specified reflected field, $\psi_\text{r}(\mathbf{r})$, and specified, transmitted field, $\psi_\text{t}(\mathbf{r})$.}
	\label{fig3_MS_GSTC}
\end{figure}

This problem can be solved by the general synthesis technique described in~\cite{GSTC,general_MS}, which is based on Generalized Sheet Transition Conditions (GSTCs) and bianisotropic surface susceptibility tensors. Figure~\ref{fig3_MS_GSTC} poses the general metasurface synthesis problem, which is to determine the surface susceptibility function $\overline{\overline{\chi}}(x,y)$ that transforms the specified incident field, $\psi_\text{i}(\mathbf{r})$, into the specified reflected field, $\psi_\text{r}(\mathbf{r})$, and specified, transmitted field, $\psi_\text{t}(\mathbf{r})$.

Here, we assume that the antenna source produces a monochromatic plane wave, with the time-harmonic phasor convention $\exp(+\mathrm{j}\omega{t})$ used throughout this paper. We set $\psi_\text{i}(\mathbf{r})$ to this field, $\psi_\text{r}(\mathbf{r})$ to be zero for matching, and $\psi_\text{t}(\mathbf{r})$ to be the desired reversed field. Moreover, we assume an axial homoanisotropic\footnote{We follow here the Greek-prefix terminology used in~\cite{general_MS}, where \emph{homo-} involves only the parameters ee and mm, \emph{hetero-} involves only the parameters em and me, and \emph{bi-} involves both the homo and hetero parameters.} metasurface, whose nonzero susceptibility components reduce to~$\chi_\text{ee}^{xx}(x,y)$, $\chi_\text{ee}^{yy}(x,y)$, $\chi_\text{mm}^{xx}(x,y)$, and $\chi_\text{mm}^{yy}(x,y)$. The corresponding susceptibility functions are then straightforwardly obtained as
\begin{subequations} \label{GSTC_relation}
	\begin{equation}
	\chi_\text{ee}^{xx}(x,y)=\frac{\Delta H_y(x,y)}{\mathrm{j}\omega\epsilon E_{x,\text{av}}(x,y)},
	\label{Eq_chi_a}
	\end{equation}
\begin{equation}
	\chi_\text{ee}^{yy}(x,y)=\frac{-\Delta H_x(x,y)}{\mathrm{j}\omega\epsilon E_{y,\text{av}}(x,y)},
	\label{Eq_chi_b}
\end{equation}			
\begin{equation}	
		\chi_\text{mm}^{xx}(x,y)=\frac{-\Delta E_y(x,y)}{\mathrm{j}\omega\mu H_{x,\text{av}}(x,y)},
		\label{Eq_chi_c}	
		\end{equation}
	\begin{equation}	
	\chi_\text{mm}^{yy}(x,z)=\frac{\Delta E_x(x,y)}{\mathrm{j}\omega\mu H_{y,\text{av}}(x,y)},
	\label{Eq_chi_d}	
	\end{equation}
\end{subequations}
where
\begin{subequations}\label{eq:GSTC_synt}
\begin{equation}
\Delta\boldsymbol{\psi}=\left[\boldsymbol{\psi}^\text{tan}_\text{t}-\boldsymbol{\psi}^\text{tan}_\text{i}\right]_{z=0}
\end{equation}
and
\begin{equation}
\boldsymbol{\psi}_{\text{av}}=\left[\frac{(\psi_\text{i}^\text{tan}+\psi_\text{t}^\text{tan})}{2}\right]_{z=0},
\end{equation}
with the subscript ``tan'' denoting the tangent part, and with
\begin{equation}
\boldsymbol{\psi}=\mathbf{E},\mathbf{H},
\end{equation}
\end{subequations}
which are to be obtained numerically from the first and second steps of the time-reversal procedure (Sec.~\ref{sec:TR_Foc_Des}).

The metasurface is nonuniform ($\overline{\overline{\chi}}=\overline{\overline{\chi}}(x,y)$), reciprocal ($\overline{\overline{\chi}}_\text{ee,mm}=\overline{\overline{\chi}}_\text{ee,mm}^T$) and nongyrotropic ($\chi_\text{ee,mm}^{xy,yx}=0$ with $\overline{\overline{\chi}}_\text{em,me}=0$)~\cite{GSTC,general_MS}.

\section{Hyperthermia Modeling} \label{sec:hype_model}

\subsection{Temperature Requirement}\label{subsec:temp_req}
As other noninvasive therapeutic treatments, hyperthermia must meet the challenging requirement of operating in a sufficiently intense fashion at the target point without harmfully affecting the surroundings of that point. This translates into a \emph{temperature} requirement. Specifically, the treatment of tumors requires the application of a temperature ranging from 41$^\circ$ to 45$^\circ$ for 15 to 60 minutes to  the tumor for making its cells adequately sensitive to radiation therapy or chemotherapy~\cite{intro-hyperthermia5}. At the same time, the maximum temperature allowed in the healthy surrounding tissues, according to medical safety guidelines~\cite{FDA_limit,Commission_limit}, is of 39$^\circ$, which means that the temperature rise above the standard body temperature must not exceed 2 degrees to guarantee the safety of the other parts of the brain. In summary, the minimal required temperature in the tumor is of 41-42$^\circ$, while the maximum allowed temperature in the rest of the brain is of 39$^\circ$. Since the related temperature challenge is mainly to achieve such a \emph{contrast} between the tumor and the surrounding healthy tissues, we introduce here the \emph{temperature figure-of-merit}
\begin{equation} \label{Figure-of-Merit_T}
\text{FoM}_T=\frac{T_{\text{healthy}}^{\text{max}}}{T_{\text{tumor}}^{\text{av}}},	
\end{equation}
where $T_{\text{healthy}}^{\text{max}}$ is the maximum temperature generated within the healthy tissues\footnote{The maximum temperatures around the tumor do not necessarily occur in the direct surrounding of the tumor, but also in the periphery of the head, as will be seen later.} and $T_{\text{healthy}}^{\text{av}}$ is the average temperature obtained within the tumor.

\subsection{Electromagnetic Model of the Head}\label{sec:head_mod}
\begin{figure} [h!]
\centering

\includegraphics[width=\columnwidth]{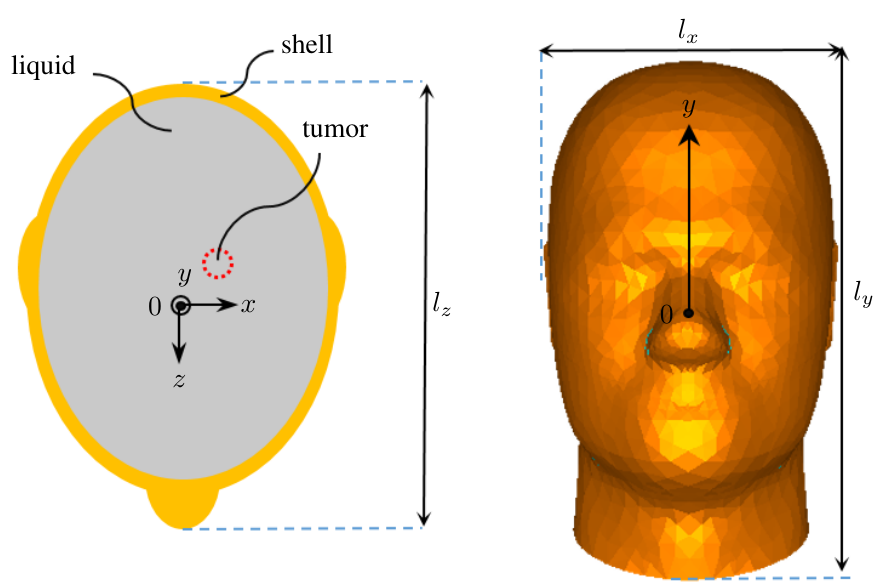}
\caption{Model of the head used in this study, consisting of the cranial shell filled by a lossy liquid that represents the brain tissues and having the shape defined in~\cite{Electrical-model1,Electrical-model2} with the dimensions $l_x=17$~cm, $l_y=29$~cm and $l_z=24$~cm. The tumor may be located anywhere within the brain. We use a coordinate system whose origin, $0$, is placed at the center of the head and that is oriented as indicated in the figure.}
	\label{fig4_headmodel}
\end{figure}

Figure~\ref{fig4_headmodel} shows the electromagnetic model of the head that is used for the numerical simulations in this study (see Sec.~\ref{sec:TR_Foc_Des}). In this model, the head is described by a two-layered voluminal structure of standard human head shape~\cite{SAM} where the outer layer corresponds to the shell and the inner layer corresponds to the brain itself. Both the shell, which is essentially constituted of bone material, and the brain, which is in reality constituted of a complex set of tissues, are approximated as homogeneous media. Table~\ref{Tab_1} gives the relative permittivity and effective conductivity of these media, as extracted at 2~GHz in~\cite{Electrical-model1,Electrical-model2}. Note that the brain has a much larger permittivity than the shell, and that it is also more conductive. Finally, the tumor will be modeled by a 5~mm radius sphere being located at arbitrary positions within the brain.

\begin{table}[h!]
		\footnotesize
		\setlength{\extrarowheight}{3pt}
		\centering
	\caption{Electrical Properties \\ of Tissues at 2~GHz \cite{Electrical-model1,Electrical-model2}}
	\begin{tabular}{lcc}
        \hline
		\textbf{Material}	& $\epsilon_r$   & $\sigma_\text{eff}\: \text{(S/m)}$ \\
		\hline\hline
		Shell & 4.7 & 0.41\\
		Brain$^\ast$ & 43  & 1.32\\ [1ex]
		\hline
	\end{tabular}
		\begin{tablenotes}
			\item$^\ast$averaged values for white matter, gray matter, fat, blood and  cerebrospinal fluid (CSF).
		\end{tablenotes}
	\label{Tab_1}
\end{table}

\subsection{Pennes Bioheat Transfer (PBT) Equation}

Our goal, according to Sec.~\ref{subsec:temp_req}, is to produce a temperature contrast with sufficiently low $\text{FoM}_T$ [Eq.~\eqref{Figure-of-Merit_T}] (within the interval mentioned) between the tumor and the rest of the head. For the modeling media and structural shape described in Sec.~\ref{sec:head_mod}, the temperature at any position within the head, $T(\mathbf{r})$, can be found by solving the Pennes bioheat transfer (PBT) equation~\cite{Bioheat-eq},
\begin{equation} \label{Eq-Bioheat}
\begin{split}
C_{p}(\mathbf{r})\rho(\mathbf{r})\frac{\partial T(\mathbf{r})}{\partial t} =& \nabla\cdot\left[K(\mathbf{r})\nabla T(\mathbf{r})\right]+A_0 (\mathbf{r})+Q(\mathbf{r}) \\
&-B(\mathbf{r})\left[T(\mathbf{r})-T_B\right],
\end{split}
\end{equation}
where $C_{p}$, $\rho$, $K$, $A_{0}$, $B$, $T_{B}$, and $Q$ are the specific heat, mass density, thermal conductivity, metabolic heat generation, capillary blood perfusion coefficient, blood temperature, and power dissipated per unit volume, respectively. This equation involves the two major mechanisms of heat transfer in such a system, viz., conduction, due temperature gradients, and convection due perfusing blood, along with the effect of distributed power dissipation. Its solution within the tumor and at the hottest point of the surrounding tissues allow to compute $\text{FoM}_T$ and hence to assess whether the hyperthermia temperature requirement is satisfied.

\subsection{Power Density}\label{sec:power_dens}

In the proposed metasurface focusing system, the power  dissipated per unit volume in the head, or power density i.e., $Q(\mathbf{r})$ in Eq.~\eqref{Eq-Bioheat}, can be directly computed from the electric field distribution, $\mathbf{E}(\mathbf{r})$, obtained by full-wave simulation, as~\cite{Time-reversal_for_HT3}
\begin{equation} \label{EQ_Q_eff}
Q(\mathbf{r})=\frac{1}{2} \sigma_\text{eff}(\mathbf{r}) |\mathbf{E}(\mathbf{r})|^2,
\end{equation}
where $\sigma_\text{eff}$ is the effective conductivity given in Tab.~\ref{Tab_1}. Given the highly lossy nature of the head structure, $Q(\mathbf{r})$ plays a dominant role in the PBT equation, as will be seen later, and can therefore be used as a proxy for $\text{FoM}_T$, and hence alleviates the systematic resolution of the differential equation. Therefore, we define here the \emph{power density figure-of-merit}
\begin{equation} \label{Figure-of-Merit_Q}
\text{FoM}_Q=\frac{Q_{\text{healthy}}^{\text{max}}}{Q_{\text{tumor}}^{\text{av}}},
\end{equation}
where $Q_{\text{healthy}}^{\text{max}}$ is the maximum power density within the healthy tissues and $Q_{\text{tumor}}^{\text{av}}$ is the average power density in the tumor.

\section{System Design} \label{sec:sys-design}

\subsection{Forward-Propagation Simulation}

\begin{figure} [h!]
\centering
\includegraphics[width=\columnwidth]{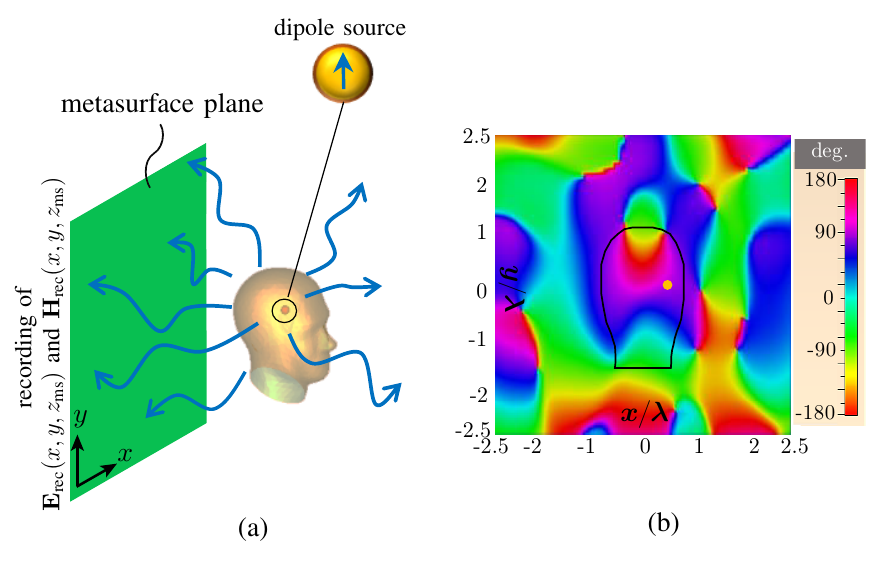}
\caption{Forward-propagation simulation step. (a)~Computation and recording of the fields scattered from  the dipole across the head in the $z_\text{ms}$-plane, $\mathbf{E}_\text{rec}(x,y,z_{\text{ms}})$ and $\textbf{H}_\text{rec}(x,y,z_{\text{ms}})$. (b)~Phase of
$E_{\text{rec},y}(x,y,z_{\text{ms}})$.}
	\label{fig5_FwSim}
\end{figure}

Following the time-reversal procedure outlined in Sec.~\ref{sec:TR_Foc_Des}, we start the design of the metasurface system by the forward-propagation simulation step [see Fig.~\ref{fig2_TR-principle}(a)], which is represented in Fig.~\ref{fig5_FwSim}.

The procedure is recalled and detailed in Fig.~\ref{fig5_FwSim}(a). We place a small electric dipole source at the location of the tumor [step 1.i)], full-wave simulate the electromagnetic field radiated from this dipole through the head, using the model of Sec.~\ref{sec:head_mod} [step 1.ii)], and record the field scattered on the plane where the metasurface will next be placed in the hyperthermia system, viz., $\mathbf{E}_\text{rec}(x,y,z_\text{ms})$ and $\mathbf{H}_\text{rec}(x,y,z_\text{ms})$ [step 1.iii)].

In all the forthcoming results, we consider a $y$-oriented dipole\footnote{This choice is essentially arbitrary. Comparable results were found for other orientations. A more sophisticated multi-dipolar or multipolar modeling, as discussed in Footnote~\ref{fn:multipole_mod}, would be appropriate for large tumors, but this is beyond the scope of this paper.}, arbitrarily positioned at the point $(x,y,z)=(10,0,-10)$~cm, the center of the 5~mm-radius volume of the tumor (see Fig.~\ref{fig4_headmodel}), and fed by a monochromatic time-harmonic source of 2~GHz. Moreover, we consider a metasurface of dimensions $5\lambda_0\times5\lambda_0$, where $\lambda_0=15$~cm is the free-space wavelength, and positioned at $z_\text{ms}=-2.5\lambda_0=-37.5$~cm. Finally, we use the commercial software CST Microwave Studio for full-wave computing the scattered fields. Figure~\ref{fig5_FwSim}(b) plots the phase of the field component $E_y$ (parallel to the dipole) in the metasurface plane for these parameters.

\subsection{Time Reversal or Electromagnetic Phase Conjugation}

We now proceed to the first part of the second step of the time-reversal procedure outlined in Sec.~\ref{sec:TR_Foc_Des} [step 2.i)], namely, the time reversal or electromagnetic phase conjugation of the field recorded in the plane of the metasurface [see top of Fig.~\ref{fig2_TR-principle}(b)].

Assuming that the fields are expressed in terms of phasors, as is the case in our full-wave simulator, the reversed  fields in the $z_\text{ms}$-plane are found as
\begin{subequations}\label{TR_design}
\begin{align}\label{TR_design_E}
\mathbf{E}_\text{rev}(x,y,z_\mathrm{ms})
&=\tilde{\mathcal{T}}\mathbf{E}_\text{rec}(x,y,z_\mathrm{ms}) \nonumber \\
&=\mathbf{E}_\text{rec}^{\ast}(x,y,z_\mathrm{ms})
\end{align}
and
\begin{align}\label{TR_design_H}
\mathbf{H}_\text{rev}(x,y,z_\mathrm{ms})
&=\tilde{\mathcal{T}}\mathbf{H}_\text{rec}(x,y,z_\mathrm{ms}) \nonumber \\
&=-\mathbf{H}_\text{rec}^{\ast}(x,y,z_\mathrm{ms}),
\end{align}
\end{subequations}
where the $\ast$ subscript represents the usual phase conjugation operation, and where the negative sign for the magnetic field is due to the time-reversal odd nature of that field (see Footnote~\ref{fn:tr}).

\begin{figure}[h!]
	\centering
\includegraphics[width=\columnwidth]{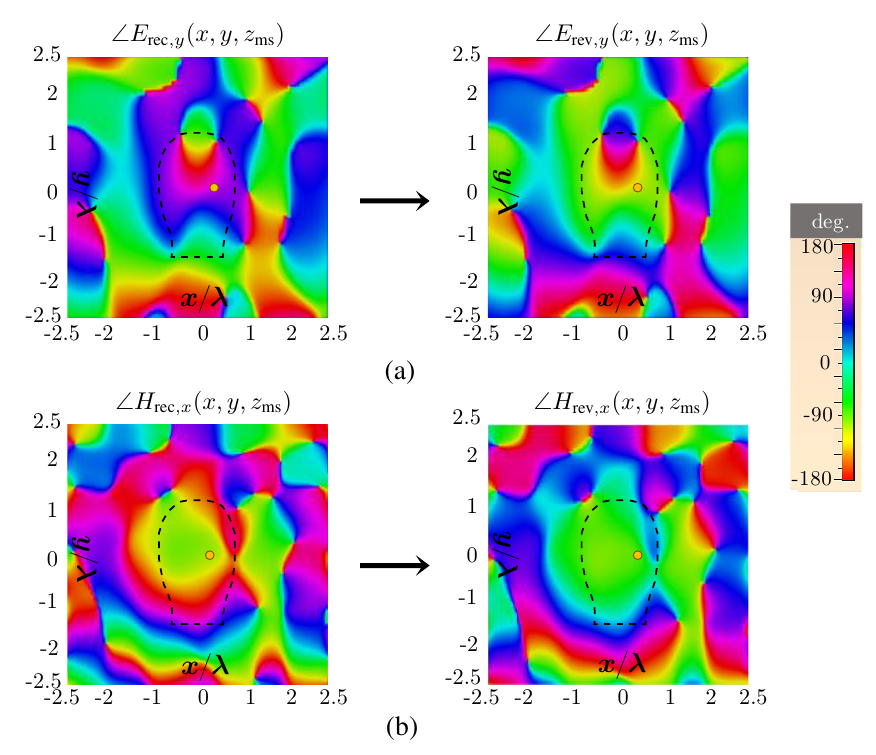}
	\caption{Time reversal or electromagnetic phase conjugation of the fields recorded in Fig.~\ref{fig5_FwSim}. (a)~$E_y$, with left panel recalled from Fig.~\ref{fig5_FwSim}(b) for convenience [Eq.~\eqref{TR_design_E}]. (b)~$H_x$  [Eq.~\eqref{TR_design_H}].}
	\label{fig6_PhCnj}
\end{figure}

Figure~\ref{fig6_PhCnj} shows the time reversal or electromagnetic phase conjugation operation, with electric field and magnetic field phase distributions plotted in Figs.~\ref{fig6_PhCnj}(a) and~\ref{fig6_PhCnj}(b), respectively, for the fields recorded in Fig.~\ref{fig5_FwSim}. The phase conjugation of the electric field trivially corresponds to the inversion of the phase, i.e., $\phi\rightarrow-\phi$, whereas the ``electromagnetic'' phase conjugation of the magnetic field, due to the time-reversal negative sign in~\eqref{TR_design_H}, implies a phase transformation of $\pi-\phi$\footnote{Considering a general complex number, $z=\rho\textrm{e}^{j\phi}$, $z^*=\rho\textrm{e}^{-j\phi}$, but $-z^*=-\rho\textrm{e}^{-j\phi}=\rho\textrm{e}^{-j\phi}\textrm{e}^{j\pi}=\textrm{e}^{j(\pi-\phi)}$.}.

\subsection{Metasurface Susceptibility Synthesis}

\begin{figure}[h!]
	\centering
\includegraphics[width=\columnwidth]{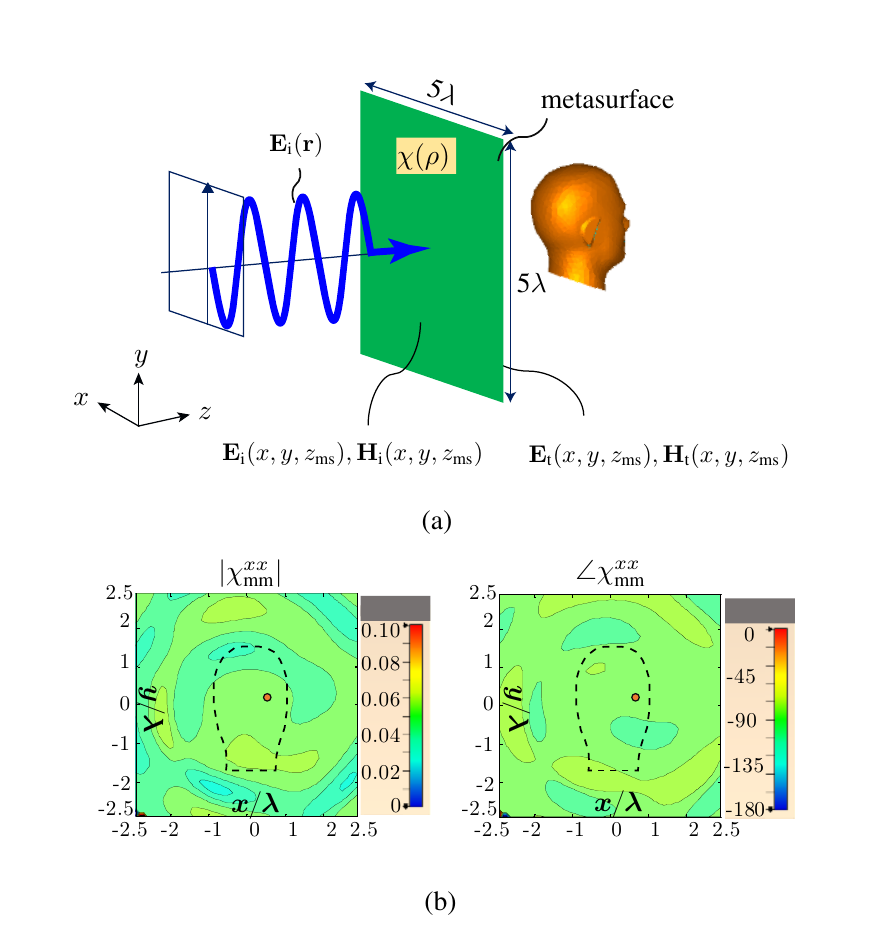}
	\caption{Metasurface susceptibility synthesis. (a)~Transformation of the field produced by the antenna source, $\mathbf{E}_\text{i}(\mathbf{r})$, into the required reversed/conjugated fields at the output plane of the metasurface, $\mathbf{E}_\text{rev}(\mathbf{r})$, given in Eq.~\eqref{fig5_FwSim}. (b)~Corresponding susceptibility function $\chi_\text{mm}^{xx}$  obtained by GSTC synthesis [Eqs.~\eqref{GSTC_relation} and~\eqref{eq:GSTC_synt}], with head projection at the center.}
	\label{fig6_illumination}
\end{figure}

With the reversed/electromagnetic-conjugated fields at our disposal, we can move on to the second part of the second step of the time-reversal procedure outlined in Sec.~\ref{sec:TR_Foc_Des} [step 2.ii)], namely, the computation of the metasurface susceptibility function transforming the field produced by the source antenna (see Fig.~\ref{fig1_metasurface system}) into the reversed/conjugated fields computed in the previous step using the GSTC technique [see bottom of Fig.~\ref{fig2_TR-principle}(b)]. This operation is represented in Fig.~\ref{fig6_illumination}.

Figure~\ref{fig6_illumination} shows the incident wave -- metasurface -- head system. The incident wave, which is typically produced by a horn antenna, is assumed to produce, in the neighbourhood of the metasurface, the monochromatic linearly $y$-polarized $+z$-propagating plane wave
\begin{align}
    \mathbf{E}_\text{i}(\mathbf{r})
    &=E_0\mathrm{e}^{j(\omega{t}-k_0z)}\hat{\mathbf{y}}, \\
    \mathbf{H}_\text{i}(\mathbf{r})
    &=-\frac{E_0}{\eta_0}\mathrm{e}^{j(\omega{t}-k_0z)}\hat{\mathbf{x}},
\end{align}
where $E_0$ is the amplitude of the electric field, which must be taken large enough to produce the required level of heat in hyperthermia, $k_0=\omega/\mathrm{c}$ is the free-space wavenumber, with \mbox{$\mathrm{c}=1/\sqrt{\epsilon_0\mu_0}$} being the speed of light in free space, and $\eta_0=\sqrt{\mu_0/\epsilon_0}=376$~$\upOmega$ is the free-space intrinsic impedance.

The field differences and averages at $z=z_\mathrm{ms}$ involved in the metasurface synthesis, assuming matching (zero reflected fields), are then found, according to Sec.~\ref{sec:ms_susc_synth}, as
\begin{subequations}
\begin{align}
\Delta E_x &= E_{x,\mathrm{rec}}^*, \\
\Delta E_y &= E_{y,\mathrm{rec}}^* - E_{y,\mathrm{i}}, \\
\Delta H_x &= -H_{x,\mathrm{rec}}^*- H_{x,\mathrm{i}}, \\
\Delta H_y &= -H_{y,\mathrm{rec}}^*, \\
E_{x,\mathrm{av}} &= E_{x,\mathrm{rec}}^*/2, \\
E_{y,\mathrm{av}} &= (E_{y,\mathrm{rec}}^* + E_{y,\mathrm{i}})/2, \\
H_{x,\mathrm{av}} &=(-H_{x,\mathrm{rec}}^*+H_{x,\mathrm{i}})/2, \\
H_{y,\mathrm{av}} &= -H_{y,\mathrm{rec}}^*/2.
\end{align}
\end{subequations}
Inserting these field specifications into the GSTC synthesis formulas in Eq.~\eqref{GSTC_relation} provides then the metasurface susceptibility functions, namely $\chi_\text{ee}^{xx}(x,y)$, $\chi_\text{ee}^{yy}(x,y)$, $\chi_\text{mm}^{xx}(x,y)$, and $\chi_\text{mm}^{yy}(x,y)$, from which the actual metasurface can be realized using standard design techniques~\cite{general_MS}. Figure~\ref{fig6_illumination} plots the magnitude and phase of the susceptibility function $\chi_\text{mm}^{xx}(x,y)$. Having comparable behaviors, the other components -- $\chi_\text{ee}^{xx}(x,y)$, $\chi_\text{ee}^{yy}(x,y)$ and $\chi_\text{mm}^{yy}(x,y)$ -- are not shown here. We see that the spatial variations of the required function $\chi_\text{mm}^{xx}(x,y)$ are  relatively slow with respect to the wavelength, indicating that the corresponding metasurface should be easy to realize with metaparticles of the typical size of about $\lambda/5$~\cite{general_MS}.

\section{Results and Discussions} \label{sec:results}

\subsection{Metasurface-Only System}
\begin{figure}[h!]
	\centering
\includegraphics[width=\columnwidth]{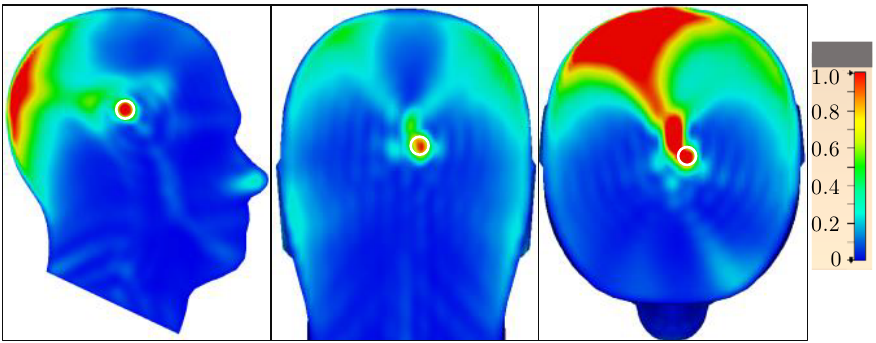}
	\caption{Normalized power density distribution, $Q(\mathbf{r})$ (W/m$^3$), for the design corresponding to Fig.~\ref{fig6_illumination} in different cross sections of the head centered on the tumor: sagittal (left), coronal (center), and axial (right). $\mathrm{FoM}_Q=2.010$. The tumor region is marked by a white circle.}
	\label{fig8_simpleMSres}
\end{figure}

Figure~\ref{fig8_simpleMSres} plots the power density distribution (Sec.~\ref{sec:power_dens}) in the three standard cross-sections -- sagittal ($yz$-plane), coronal ($xy$-plane), and axial ($zx$-plane) -- of the head centered on the tumor (see Fig.~\ref{fig4_headmodel}). Fairly good focusing is achieved in the coronal section. In contrast, focusing is poor in the sagittal and axial sections, where we observe power density spots that are as strong in some healthy regions as at the point of the tumor. This design corresponds to $\mathrm{FoM}_Q=2.010$, which means that the maximum power density in the healthy tissues is twice the average power density in the tumor.

The poorness of this result is not surprising considering that perfect focusing would require a complete time-reversing enclosure, according to the comment near the end of Sec.~\ref{sec:TR_Foc_Des}, a condition that is far from being satisfied with the single-face (only 1/6 of the cubic enclosure) used here (see Fig.~\ref{fig6_illumination}).

\subsection{Addition of Enclosing Reflecting Walls}
While a quasi-complete, 5-face time-reversing enclosure would certainly result in a quasi-perfect focusing result, such a design appears prohibitively complex, as it would also imply 5 illuminating antennas and a fairly heavy supporting structure. A more reasonable approach to improve the focusing performance, consists, as suggested at the end of Sec.~\ref{sec:TR_Foc_Des}, to enrich the scattering diversity around the region of interest. One way to accomplish this is to add enclosing metallic reflecting walls around the targeted focal region~\cite{TR-Cavity3}. Indeed, even if they do not perform time reversal, such walls reflect the part of the wave that would otherwise be lost in space and direct partly back to the time-reversing wall which can then make it contribute to focusing after reflection. We leverage here this strategy by adding 2 to 4 metallic walls around the head.

\begin{figure}[h!]
	\centering
\includegraphics[width=\columnwidth]{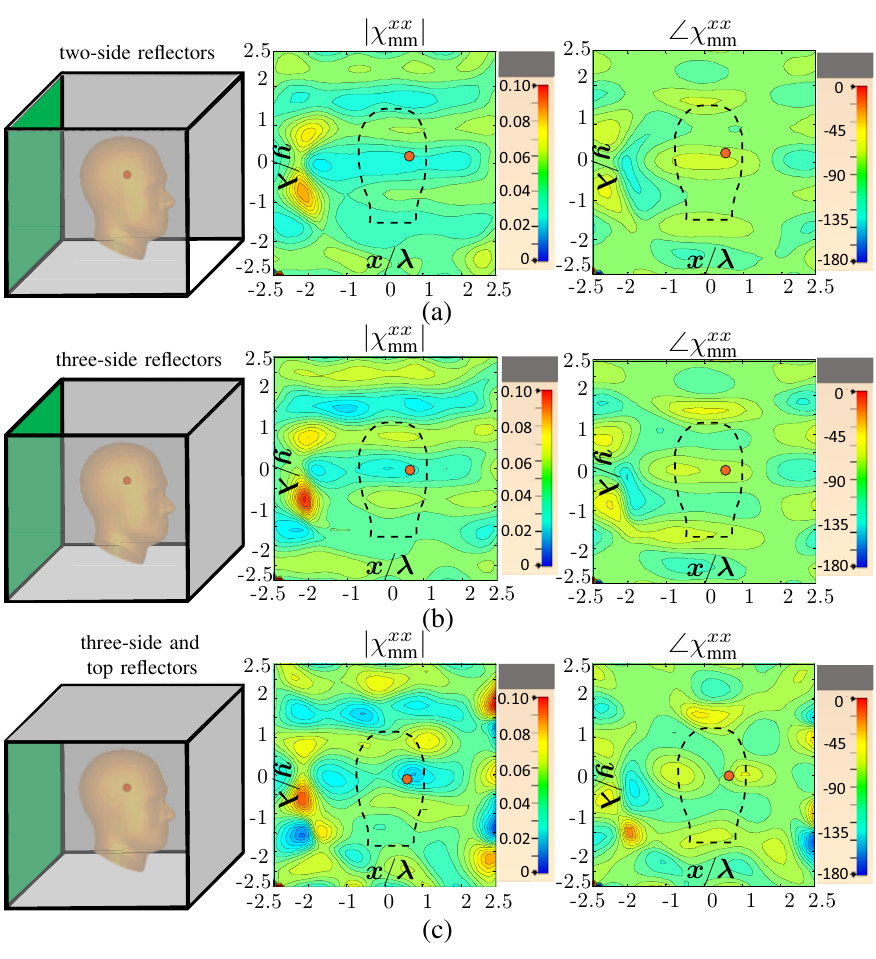}
	\caption{Addition of enclosing (PEC) reflecting walls to the metasurface-only setup of Fig.~\ref{fig6_illumination} and corresponding susceptibility functions $\chi_\text{mm}^{xx}(x,y)$. (a)~Walls at the two sides of the head. (b)~Three walls around the head. (c)~Same as (b) plus wall at the top of the head.}
	\label{fig9_pec+ms_Chi}
\end{figure}

Figure~\ref{fig9_pec+ms_Chi} shows three setups with added enclosing PEC reflecting walls and plots the susceptibility functions obtained for these new setups. These functions are considerably different from that for the metasurface-only system in Fig.~\ref{fig6_illumination} and also between each other, which suggest that the reflecting walls may play a considerable role in the operation of the system.

\begin{figure}[h!]
\centering
		\centering
		\includegraphics[width=\columnwidth]{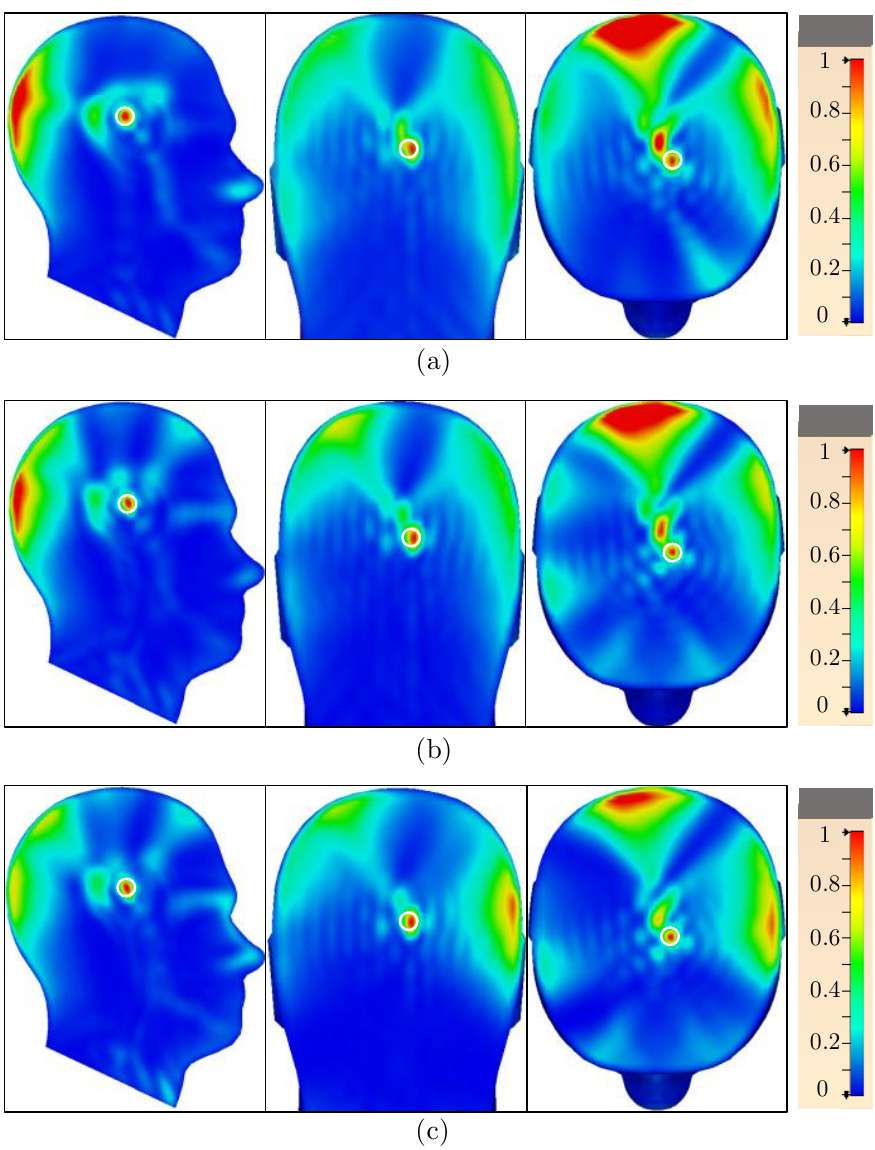}
		\caption{Effect, in normalized power density, of adding enclosing (PEC) reflecting walls around the head according to the setups at the left of Fig.~\ref{fig9_pec+ms_Chi}(a)-(c). (a)~Walls at the two sides, $\text{FoM}_Q=1.646$. (b)~Walls all around, $\text{FoM}_Q=1.438$. (c)~Same as (b) plus top, $\text{FoM}_Q=1.048$.}
	\label{fig10_pec+ms_Q}

\end{figure}

Figure~\ref{fig10_pec+ms_Q} plots the focusing results for the setups of Fig.~\ref{fig9_pec+ms_Chi}. These results confirm the anticipated significant effect of adding enclosing reflecting walls to the metasurface. We observe indeed a global improvement of the focusing performance compared to the result for the metasurface-only setup in Fig.~\ref{fig8_simpleMSres}, as well as a progressive improvement when increasing the number of conducting walls, especially in the sagittal and axial sections. This is confirmed by the monotonically decreasing $\text{FoM}_Q$, from $2.010$ in the metasurface-only setup to $\text{FoM}_Q=1.048$ in the 4-wall setup. In the 4-wall setup, the undesirable power concentration spots in the healthy parts of the brain have been considerably reduced. However, some hot spots remain at the edges of the cranial cavity, and the fact that $\text{FoM}_Q$ is still larger than one, even in the best case ($\text{FoM}_Q=1.048$), indicates that the maximum power density in the healthy tissues is still larger than the average power density in the tumor. Therefore, further improvement of the system is required.

\subsection{Addition of a Scatterer}
\begin{figure}[h!]
	\centering
\includegraphics[width=\columnwidth]{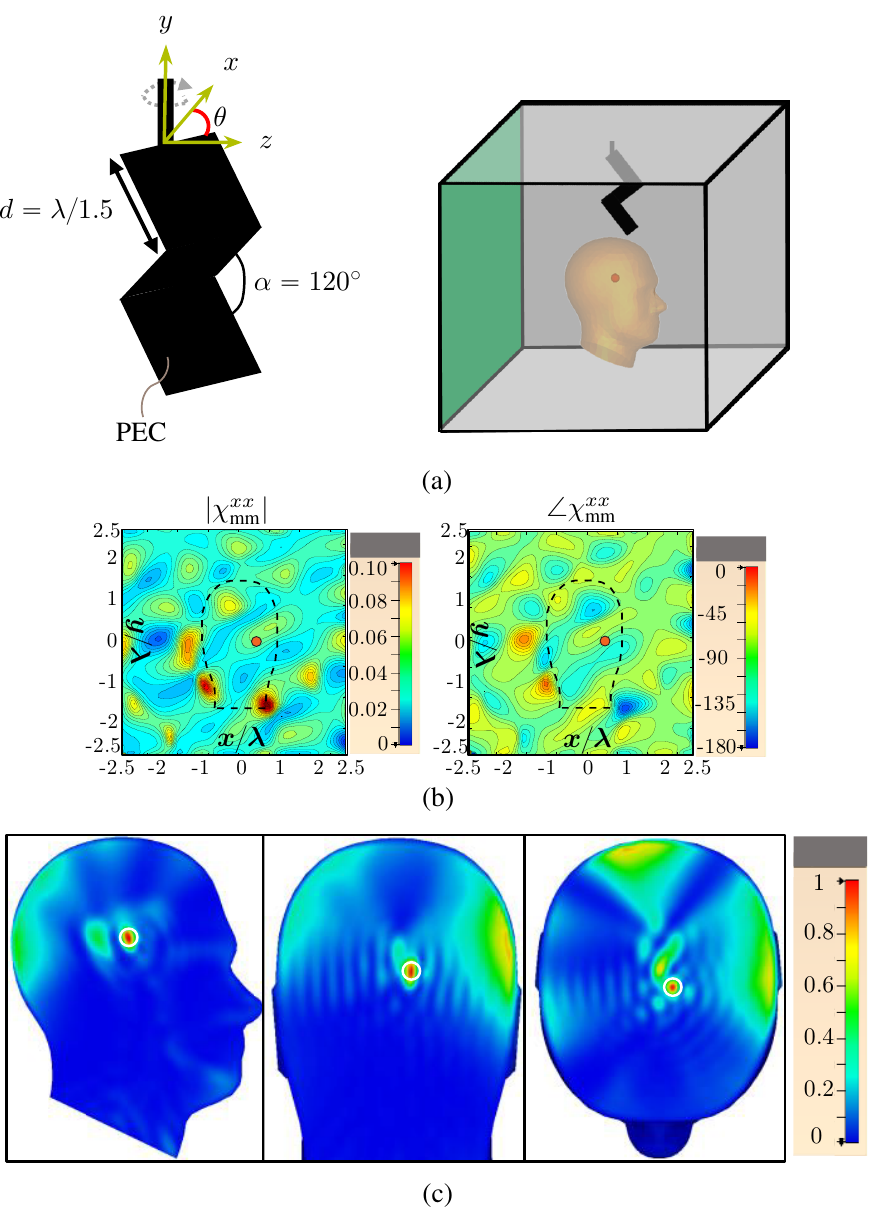}
	\caption{Study of the effect of the addition of a scatterer to the setup of Fig.~\ref{fig9_pec+ms_Chi}(c). (a)~The scatterer, and placement above the head of the patient in the cavity. (b)~Corresponding susceptibility function $\chi_\text{mm}^{xx}(x,y)$. (c)~Normalized power density distribution in the sagittal (left), coronal (center), and axial (right) tumor-centered cross-sections of the head. $\text{FoM}_Q=0.617$.}
	\label{fig11_scatterer}
\end{figure}

We shall now leverage an extra technique that has proven effective for enhancing the quality of time-reversal focusing: inserting a strong scatterer in the proximity of the targeted focus within the cavity of the system~\cite{Mode-stirrer1,Mode-stirrer2,Mode-stirrer3}. Such a scatterer acts as a cavity mode transformer that can be optimized to favourably redirect part of the incoming wave, which could otherwise be lost, toward the target\footnote{This might be easiest to understand with a forward-scattering thought experiment: if the scatterer is properly designed, the wave radiated by the dipole through the structure of interest (here the head), can get favourably redirected by the scatterer toward the time-reversing active wall, where it will provide optimal extra spatial information for focusing~\cite{Scatterer-TR-focusing}.}. Our related study is presented in Fig.~\ref{fig11_scatterer}.

We add here a scatterer to the best PEC-wall setup of the previous section, i.e., the setup with 4 reflecting walls. We choose a scatterer consisting of a twice zigzag-folded rectangular PEC plate and place it above the head of the patient, as shown in Fig.~\ref{fig11_scatterer}(a). Figure~\ref{fig11_scatterer}(b) shows the corresponding susceptibility function. We note that this function is quite different from the scatterer-less function in Fig.~\ref{fig9_pec+ms_Chi}(c), which let us expect a substantial improvement. This expectation is confirmed in the power density distribution result, given in Fig.~\ref{fig11_scatterer}(c), where the hot spots in the healthy tissues have almost completely disappeared and where the $\text{FoM}_Q=0.617$ indicates that the maximum power density in the healthy tissues has now reduced to about 60$\%$ of the average power density in the tumor. This represents an improvement of about $40\%$ compared to the design of Fig.~\ref{fig9_pec+ms_Chi}(c).

\begin{figure}[h!]
	\centering
\includegraphics[width=\columnwidth]{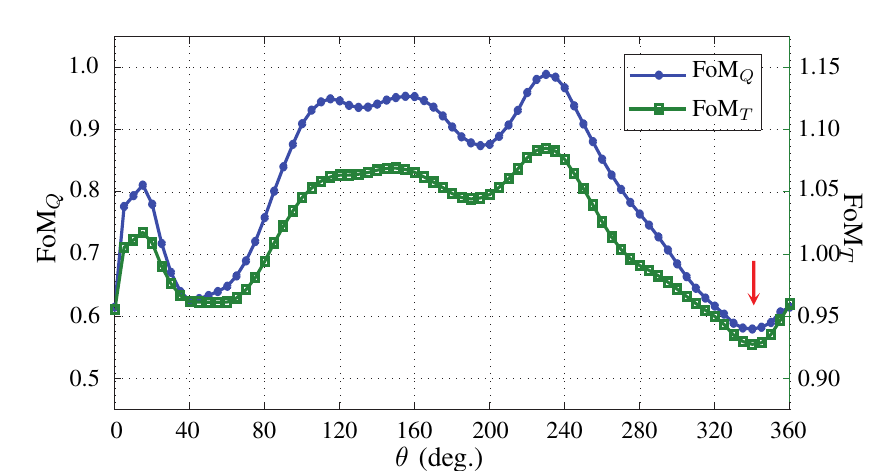}
	\caption{Figures of merit $\text{FoM}_Q$ [Eq.~\eqref{Figure-of-Merit_Q}] and $\text{FoM}_T$ [Eq.~\eqref{Figure-of-Merit_T}] versus orientation angle, $\theta$, in Fig.~\ref{fig11_scatterer}(a). The arrow indicates the optimal angle, $\theta_\mathrm{opt}=340^\circ$, where $\text{FoM}_Q=0.578$ and $\text{FoM}_T=0.927$. Figures~\ref{fig11_scatterer}(b) and~(c) correspond to $\theta=0$.}
	\label{fig12_rot}
\end{figure}

\begin{figure}[h!]
		\centering
\includegraphics[width=\columnwidth]{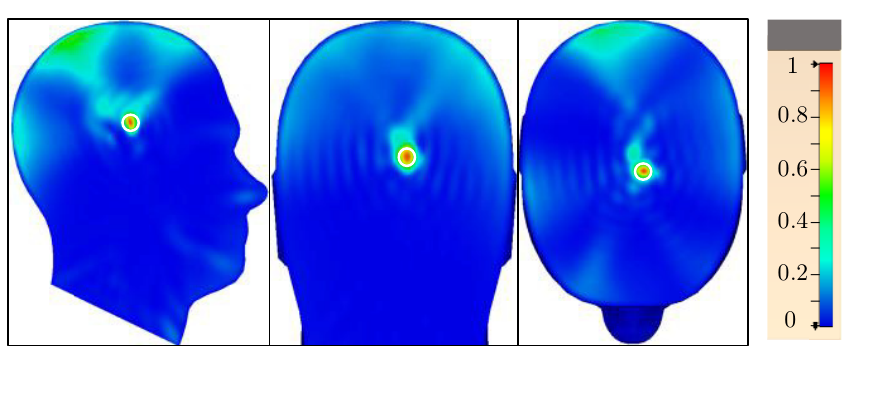}
\caption{Normalized power density distributions for the wall-scatter setup of Fig.~\ref{fig11_scatterer} for the optimal scatter orientation found in Fig.~\ref{fig12_rot} ($\theta_\mathrm{opt}=340^\circ$).}
	\label{fig13_idealQ}
\end{figure}

In the setup of Fig.~\ref{fig11_scatterer}, the scatterer was oriented in an arbitrary fashion. We shall now study the effect of the orientation of this scatterer by varying the angle $\theta$ in Fig.~\ref{fig11_scatterer}. The corresponding results are plotted in terms of $\text{FoM}_Q$ and $\text{FoM}_T$ versus $\theta$ in Fig.~\ref{fig12_rot}, where we incidentally note that the $\text{FoM}_Q$, directly obtainable from the computed electric field via Eq.~\eqref{EQ_Q_eff}, is here a fairly good proxy for $\text{FoM}_T$, requiring the resolution of the PBT equation [Eq.~\eqref{Eq-Bioheat}]. This result reveals that the system is strongly sensitive to the orientation of the scatterer, with a $\text{FoM}_T$ variation of around $17\%$. The lowest level of both $\text{FoM}_Q$ and $\text{FoM}_T$, and hence the optimal orientation of the scatterer, is found to occur at $\theta=340^\circ$, where $\text{FoM}_Q=0.578$ and $\text{FoM}_T=0.927$. The improvement is even apparent by comparing the power density distributions in Figs.~\ref{fig13_idealQ} and those in Fig.~\ref{fig11_scatterer}(c), where the strongest undesired intensity spots in the coronal and axial cross-sections have been essentially suppressed.

\subsection{Temperature Profile}

Although the power density is strongly related to the temperature in our system (see Fig.~\ref{fig12_rot}), it is ultimately the temperature quantity that matters in hyperthermia. We shall therefore examine here the exact temperature distribution in the head.
The temperature distribution can be directly computed by the commercial software (CST Microwave Studio) as a solution of the PBT equation [Eq.~\eqref{Eq-Bioheat}], which involves the power density, $Q(\mathbf{r})$, that is itself related to the intensity of the electric field distribution according to Eq.~\eqref{EQ_Q_eff}. We used in our simulation the PBT parameters provided by~\cite{Thermal_model2}, which are listed in Tab.~\ref{Tab2_Thermal}.

\begin{table}[h!]
\footnotesize
		\setlength{\extrarowheight}{3pt}
		\centering
	\caption{Thermal properties of the human head at 2~GHz.}

	\begin{tabular} {p{8mm}p{9mm}p{17mm}p{16mm}p{19mm}}
		\hline
		\textbf{Material}
		& $\rho\:(\text{kg$/$m}^{3})$
		& $K\:(\text{W$/$(m$\cdot$}^\circ \text{C}))$
		& $C_{p}\:(\text{J$/$(kg$\cdot$}^\circ \text{C}))$
		& $B\:(\text{W$/$(m}^3\cdot^\circ \text{C}))$ \\
	\hline\hline	
		Shell	& 1000  & 0.210 & 2500 & 1000		\\
		Brain	& 1030 & 0.502 & 3700 & 40000	\\ [1ex]
		\hline
	\end{tabular}
	\label{Tab2_Thermal}
\end{table}

Figure~\ref{fig13_idealQ} shows the temperature distribution reached after 20 minutes of exposure for the optimal design, corresponding to the power density distribution in Fig.~\ref{fig13_idealQ}. This simulation accounts for the typical 20$^\circ$ thin water bolus that is typically placed around to the head to prevent skin burning in clinical hyperthermia, by keeping the skin side of the cranial cavity to 20$^\circ$. This result, with temperature of $42^\circ$ in the tumor region and temperatures not exceeding $39^\circ$ in the rest of the brain, where it is  about $37^\circ$ almost everywhere, shows that the proposed system satisfies the temperature requirements in Sec.~\ref{subsec:temp_req}, and hence demonstrates its applicability to real hyperthermia treatments.

\begin{figure}[h!]
	\centering
\includegraphics[width=\columnwidth]{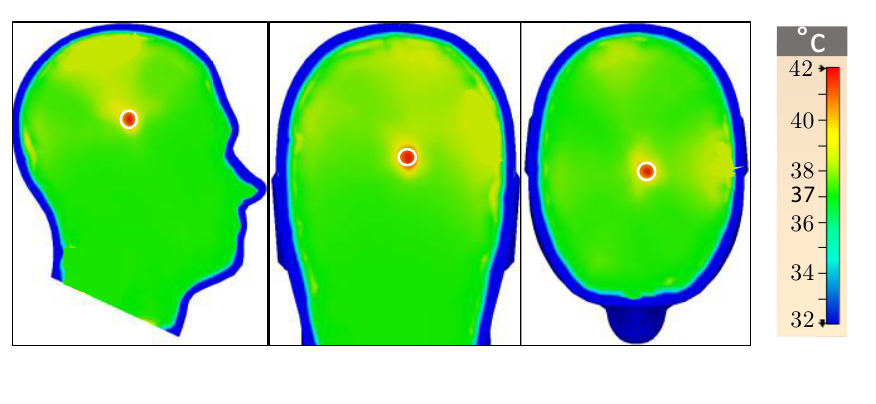}
	\caption{Temperature ($^\circ$C) distribution reached after 20 minutes of exposure for the optimal design ($Q(\mathbf{r})$ in Fig.~\ref{fig13_idealQ}) in the sagittal (left), coronal (center), and axial (right) tumor-centered cross-sections of the head.}
	\label{fig14_Temp}
\end{figure}

\section{Conclusion}  \label{sec:conclusion}
We have proposed and full-wave demonstrated a novel, metasurface-based microwave hyperthermia system for the treatment of deep-seated brain tumors, where optimal focusing is achieved by a time-reversal design of the metasurface, assisted by metallic reflecting walls and a scatterer. Compared to other hyperthermia technologies, this system allows finer sampling and therefore higher focusing resolution, in addition to featuring lower complexity and hence lower cost.

Given recent progress in metasurface technology, this system can be easily realized and optimized for applicability to clinical hypothermia treatment. Further improvements would be achieved by replacing the current flat metasurface by a metasurface that would be cylindrically~\cite{Curved_MS_Mojtaba} or spherically~\cite{Jia_TAP_04_2019} curved around the head. Moreover, the metasurface should be ideally made reconfigurable and programmable~\cite{MS-programmable,Wang_2021} for matching different head sizes and shapes in a real-life scenario.

\bibliographystyle{IEEEtran}
\bibliography{IEEEabrv, Hajiahmadi_TRMS_Hyperthermia_ref}

\begin{thebibliography}{10}
\providecommand{\url}[1]{#1}
\csname url@samestyle\endcsname
\providecommand{\newblock}{\relax}
\providecommand{\bibinfo}[2]{#2}
\providecommand{\BIBentrySTDinterwordspacing}{\spaceskip=0pt\relax}
\providecommand{\BIBentryALTinterwordstretchfactor}{4}
\providecommand{\BIBentryALTinterwordspacing}{\spaceskip=\fontdimen2\font plus
\BIBentryALTinterwordstretchfactor\fontdimen3\font minus
  \fontdimen4\font\relax}
\providecommand{\BIBforeignlanguage}[2]{{%
\expandafter\ifx\csname l@#1\endcsname\relax
\typeout{** WARNING: IEEEtran.bst: No hyphenation pattern has been}%
\typeout{** loaded for the language `#1'. Using the pattern for}%
\typeout{** the default language instead.}%
\else
\language=\csname l@#1\endcsname
\fi
#2}}
\providecommand{\BIBdecl}{\relax}
\BIBdecl

\bibitem{intro-hyperthermia1}
P.~K. Sneed~\emph{et al}, ``Survival benefit of hyperthermia in a prospective
  randomized trial of brachytherapy boost ± hyperthermia for glioblastoma
  multiforme,'' \emph{Int. J. Radiat. Oncol. Biol. Phys.}, vol.~40, no.~2, pp.
  287--295, 1998.

\bibitem{intro-hyperthermia2}
R.~B. Roemer, ``Engineering aspects of hyperthermia therapy,'' \emph{Annu. Rev.
  Biomed. Eng.}, vol.~1, no.~1, pp. 347--376, 1999.

\bibitem{intro-hyperthermia3}
J.~van~der Zee, ``Engineering aspects of hyperthermia therapy,'' \emph{Ann.
  Oncol.}, vol.~13, no.~8, pp. 1173--1184, 2002.

\bibitem{intro-hyperthermia4}
J.~R. Lepock, ``Cellular effects of hyperthermia: Relevance to the minimum dose
  for thermal damage,'' \emph{Int. J. Hyperthermia}, vol.~19, no.~3, pp.
  252--266, 2003.

\bibitem{intro-hyperthermia5}
P.~R. Stauffer, ``Evolving technology for thermal therapy of cancer,''
  \emph{Int. J. Hyperthermia}, vol.~21, no.~8, pp. 731--744, 2005.

\bibitem{Microwave_Hyperthermia1}
A.~J. Fenn, G.~L. Wolf, and R.~M. Fogle, ``An adaptive microwave phased array
  for targeted heating of deep tumours in intact breast: Animal study
  results,'' \emph{Int. J. Hyperthermia}, vol.~15, no.~1, pp. 45--61, 1999.

\bibitem{Microwave_Hyperthermia2}
M.~{Converse}, E.~J. {Bond}, B.~D. {Veen}, and C.~{Hagness}, ``A computational
  study of ultra-wideband versus narrowband microwave hyperthermia for breast
  cancer treatment,'' \emph{IEEE Trans. Microw. Theory Tech.}, vol.~54, no.~5,
  pp. 2169--2180, 2006.

\bibitem{Microwave_Hyperthermia3}
K.~{Arunachalam}, S.~S. {Udpa}, and L.~{Udpa}, ``Computational feasibility of
  deformable mirror microwave hyperthermia technique for localized breast
  tumors,'' \emph{Int. J. Hyperthermia}, vol.~23, no.~7, pp. 577--589, 2007.

\bibitem{Microwave_Hyperthermia4}
E.~{Zastrow}, S.~C. {Hagness}, and B.~D. {Van Veen}, ``3d computational study
  of non-invasive patient-specific microwave hyperthermia treatment of breast
  cancer,'' \emph{Phys. Med. Biol.}, vol.~55, pp. 3611--3629, 2010.

\bibitem{Microwave_Hyperthermia5}
H.~Kok~\emph{et al}, ``Current state of the art of regional hyperthermia
  treatment planning: a review,'' \emph{Radiation Oncology}, vol.~10, 2015.

\bibitem{Time-reversal_for_HT1}
{Bin Guo}, {Luzhou Xu}, and {Jian Li}, ``Time reversal based microwave
  hyperthermia treatment of breast cancer,'' in \emph{Asilomar Conference Sign.
  Syst. and Comp..}, Oct. 2005, pp. 290--293.

\bibitem{Time-reversal_for_HT2}
P.~{Kosmas}, E.~{Zastrow}, S.~C. {Hagness}, and B.~D. {Van Veen}, ``A
  computational study of time reversal techniques for ultra-wideband microwave
  hyperthermia treatment of breast cancer,'' in \emph{14th Workshop on
  Statistical Signal Processing (SP)}, Sept. 2007, pp. 312--316.

\bibitem{Time-reversal_for_HT4}
J.~{Stang}, M.~{Haynes}, P.~{Carson}, and M.~{Moghaddam}, ``A preclinical
  system prototype for focused microwave thermal therapy of the breast,''
  \emph{IEEE Trans. Biomed. Eng.}, vol.~59, no.~9, pp. 2431--2438, Sept. 2012.

\bibitem{TR-Ultrasonic}
M.~Fink, ``Time reversal of ultrasonic fields. i. basic principles,''
  \emph{IEEE Trans. Ultrason. Ferroelectr. Freq. Control}, vol.~39, no.~5, pp.
  555--566, Sept. 1992.

\bibitem{TR-Microwave1}
G.~Lerosey, J.~de~Rosny, A.~Tourin, A.~Derode, G.~Montaldo, and M.~Fink, ``Time
  reversal of electromagnetic waves,'' \emph{Phys. Rev. Lett.}, vol.~92,
  no.~19, p. 193904, 2004.

\bibitem{TR-definition}
J.~D. Rosny, G.~Lerosey, and M.~Fink, ``Theory of electromagnetic time-reversal
  mirrors,'' \emph{IEEE Trans. Antennas Propag.}, vol.~58, no.~10, pp.
  3139--3149, Oct. 2010.

\bibitem{TR-Microwave}
M.~Davy, J.~Minonzio, J.~de~Rosny, C.~Prada, and M.~Fink, ``Experimental study
  of the invariants of the time-reversal operator for a dielectric cylinder
  using separate transmit and receive arrays,'' \emph{IEEE Trans. Antennas
  Propag.}, vol.~58, no.~4, pp. 1349--1356, April 2010.

\bibitem{TR-definition1}
M.~Fink, C.~Prada, F.~Wu, and D.~Cassereau, ``Self focusing with time reversal
  mirror in inhomogeneous media,'' in \emph{IEEE Ultrason. Symp.}, 1989, pp.
  681--686 vol.2.

\bibitem{Time-reversal_for_HT3}
H.~D. Trefn{\`a}~\textit{et al}, ``Time-reversal focusing in microwave
  hyperthermia for deep-seated tumors,'' \emph{Phys. Med. Biol.}, vol. 55,
  2167, 2010.

\bibitem{Antenna_Complexity3}
P.~{Takook}, M.~{Persson}, and H.~D. {Trefná}, ``Performance evaluation of
  hyperthermia applicators to heat deep-seated brain tumors,'' \emph{IEEE J.
  Electromagn. RF Microw. Med. Biol.}, vol.~2, no.~1, pp. 18--24, 2018.

\bibitem{Antenna_Complexity1}
P.~T. {Nguyen}, A.~{Abbosh}, and S.~{Crozier}, ``Three-dimensional microwave
  hyperthermia for breast cancer treatment in a realistic environment using
  particle swarm optimization,'' \emph{IEEE Trans. Biomed. Eng.}, vol.~64,
  no.~6, pp. 1335--1344, 2017.

\bibitem{Antenna_Complexity2}
E.~{Zastrow}, S.~C. {Hagness}, B.~D. {Van Veen}, and J.~E. {Medow},
  ``Time-multiplexed beamforming for noninvasive microwave hyperthermia
  treatment,'' \emph{IEEE Trans. Biomed. Eng.}, vol.~58, no.~6, pp. 1574--1584,
  2011.

\bibitem{complex_feeding_network1}
Z.~{Nie} and Y.~{Yang}, ``A model independent scheme of adaptive focusing for
  wireless powering to in-body shifting medical device,'' \emph{IEEE Trans.
  Antennas Propag.}, vol.~66, no.~3, pp. 1497--1506, 2018.

\bibitem{complex_feeding_network2}
T.~{Fromenteze}, D.~{Carsenat}, and C.~{Decroze}, ``A precorrection method for
  passive uwb time-reversal beamformer,'' \emph{IEEE Antennas Wireless Propag.
  Lett.}, vol.~12, pp. 836--840, 2013.

\bibitem{GSTC}
K.~Achouri and C.~Caloz, ``Design, concepts, and applications of
  electromagnetic metasurfaces,'' \emph{Nanophotonics}, vol.~7, no.~6, pp.
  1095--1116, Jun. 2018.

\bibitem{general_MS}
------, \emph{Electromagnetic Metasurfaces: Theory and Applications}.\hskip 1em
  plus 0.5em minus 0.4em\relax Wiley-IEEE Press, 2021.

\bibitem{WFS_MS_scirep}
N.~Kaina, M.~Dupré, G.~Lerosey, and M.~Fink, ``Shaping complex microwave
  fields in reverberating media with binary tunable metasurfaces,'' \emph{Sci.
  Rep.}, vol. 4, 6693, 2014.

\bibitem{SpatialProcessor_MS}
K.~{Achouri}, G.~{Lavigne}, M.~A. {Salem}, and C.~{Caloz}, ``Metasurface
  spatial processor for electromagnetic remote control,'' \emph{IEEE Trans.
  Antennas Propag.}, vol.~64, no.~5, pp. 1759--1767, 2016.

\bibitem{THz_WFS_MS}
H.~{Yi}, S.~{Qu}, K.~{Ng}, C.~K. {Wong}, and C.~H. {Chan}, ``Terahertz
  wavefront control on both sides of the cascaded metasurfaces,'' \emph{IEEE
  Trans. Antennas Propag.}, vol.~66, no.~1, pp. 209--216, 2018.

\bibitem{WFS_MS_nature}
M.~Jang~\emph{et al}, ``Wavefront shaping with disorder-engineered
  metasurfaces,'' \emph{Nat. Photonics}, vol.~12, pp. 84--90, 2018.

\bibitem{TR-Cavity1}
D.~{Cassereau} and M.~{Fink}, ``Time-reversal of ultrasonic fields. iii. theory
  of the closed time-reversal cavity,'' \emph{IEEE Trans. Ultrason.
  Ferroelectr. Freq. Control}, vol.~39, no.~5, pp. 579--592, 1992.

\bibitem{TR-Cavity2}
R.~Carminati, R.~Pierrat, J.~de~Rosny, and M.~Fink, ``Theory of the time
  reversal cavity for electromagnetic fields,'' \emph{Opt. Lett.}, vol.~32,
  no.~21, pp. 3107--3109, 2007.

\bibitem{TR-Cavity3}
S.~{Ding}, Y.~{Fang}, J.~{Zhu}, Y.~{Yang}, and B.~{Wang}, ``Wireless cloaking
  system based on time-reversal multipath propagation effects,'' \emph{IEEE
  Trans. Antennas Propag.}, vol.~67, no.~2, pp. 1386--1391, 2019.

\bibitem{TR-Cavity4}
C.~Caloz and Z.~Deck-L\'{e}ger, ``Spacetime metamaterials—part ii: Theory and
  applications,'' \emph{IEEE Trans. Antennas Propag.}, vol.~68, no.~3, pp.
  1583--1598, 2020.

\bibitem{Scatterer-TR-focusing}
M.~E. {Yavuz} and F.~L. {Teixeira}, ``A numerical study of time-reversed uwb
  electromagnetic waves in continuous random media,'' \emph{IEEE Antennas
  Wireless Propag. Lett.}, vol.~4, pp. 43--46, 2005.

\bibitem{collin1990field}
R.~E. Collin, \emph{Field Theory of Guided Waves}.\hskip 1em plus 0.5em minus
  0.4em\relax John Wiley \& Sons, 1990, vol.~5.

\bibitem{PolarizationCurrents_Sarkar}
T.~K. {Sarkar} and E.~{Arvas}, ``An integral equation approach to the analysis
  of finite microstrip antennas: volume/surface formulation,'' \emph{IEEE
  Trans. Antennas Propag.}, vol.~38, no.~3, pp. 305--312, 1990.

\bibitem{Ishimaru_1990}
A.~Ishimaru, \emph{Electromagnetic Wave Propagation, Radiation, and Scattering:
  From Fundamentals to Applications}, 2nd~ed.\hskip 1em plus 0.5em minus
  0.4em\relax Wiley - IEEE Press, 2017.

\bibitem{Caloz_PRAp_10_2018}
C.~Caloz, A.~Al\`{u}, S.~Tretyakov, D.~Sounas, K.~Achouri, and Z.-L.
  Deck-L\'{e}ger, ``Electromagnetic nonreciprocity,'' \emph{Phys. Rev. Appl.},
  vol.~10, no.~4, pp. 047\,001:1--26, Oct. 2018, invited.

\bibitem{general_Huygens}
R.~Courant and D.~Hilbert, \emph{Methods of Mathematical Physics}.\hskip 1em
  plus 0.5em minus 0.4em\relax John Wiley \& Sons, 1962, vol.~2.

\bibitem{FDA_limit}
\BIBentryALTinterwordspacing
Magnetic {Resonance} {Imaging} {(MRI)} {Safety}. [Online]. Available:
  \url{https://www.fda.gov/radiation-emitting-products/medical-imaging/mri-magnetic-resonance-imaging}
\BIBentrySTDinterwordspacing

\bibitem{Commission_limit}
{International Commission on Non-Ionising Radiation (ICNIRP)}, ``Guidelines for
  limiting exposure for time varying electric, magnetic and electromagnetic
  fields (up to 300 {GHz}),'' \emph{Health Phys.}, vol.~74, no.~4, pp.
  494--522, 1998.

\bibitem{SAM}
C.~G. Gordon~\emph{et al}, \emph{1988 Anthropometric survey of U.S. army
  personnel: methods and summary statistics}.\hskip 1em plus 0.5em minus
  0.4em\relax Technical Report NATICK/TR‐89/044, U.S. Army Natick Research,
  Development and Engineering Center, Natick, Massachusetts, 1989.

\bibitem{Electrical-model1}
M.~Consultants, \emph{Dielectric Database.}\hskip 1em plus 0.5em minus
  0.4em\relax London, U.K.: Microwave Consultants, 1994.

\bibitem{Electrical-model2}
N.~{Joachimowicz}, B.~{Duchêne}, C.~{Conessa}, and O.~{Meyer}, ``Reference
  phantoms for microwave imaging,'' in \emph{11th European Conference on
  Antennas and Propagation (EUCAP)}, 2017, pp. 2719--2722.

\bibitem{Bioheat-eq}
H.~H. Pennes, ``Analysis of tissue and arterial blood temperatures in the
  resting human forearm,'' \emph{J. Appl. Physiol.}, vol.~85, 1998.

\bibitem{Mode-stirrer1}
P.~{Plaza-Gonzalez}, J.~{Monzo-Cabrera}, J.~M. {Catala-Civera}, and
  D.~{Sanchez-Hernandez}, ``Effect of mode-stirrer configurations on dielectric
  heating performance in multimode microwave applicators,'' \emph{IEEE Trans.
  Microw. Theory Tech.}, vol.~53, no.~5, pp. 1699--1706, 2005.

\bibitem{Mode-stirrer2}
L.~Arnaut, ``Mode-stirred reverberation chambers: A paradigm for
  spatio-temporal complexity in dynamic electromagnetic environments,''
  \emph{Wave Motion}, vol.~51, no.~4, pp. 673--684, 2014.

\bibitem{Mode-stirrer3}
H.~Sun~\emph{et al}, ``Parametric testing of metasurface stirrers for
  metasurfaced reverberation chambers,'' \emph{Sensors}, vol. 19(4):976, 2019.

\bibitem{Thermal_model2}
G.~T. {Martin}, M.~G. {Haddad}, E.~G. {Cravalho}, and H.~F. {Bowman}, ``Thermal
  model for the local microwave hyperthermia treatment of benign prostatic
  hyperplasia,'' \emph{IEEE Trans. Biomed. Eng.}, vol.~39, no.~8, pp. 836--844,
  1992.

\bibitem{Curved_MS_Mojtaba}
M.~{Dehmollaian}, N.~{Chamanara}, and C.~{Caloz}, ``Wave scattering by a
  cylindrical metasurface cavity of arbitrary cross section: Theory and
  applications,'' \emph{IEEE Trans. Antennas Propag.}, vol.~67, no.~6, pp.
  4059--4072, 2019.

\bibitem{Jia_TAP_04_2019}
X.~Jia, Y.~Vahabzadeh, C.~Caloz, and F.~Yan, ``Synthesis of spherical
  metasurfaces based on susceptibility tensor {GSTC}s,'' \emph{IEEE Trans.
  Antennas Propag.}, vol.~67, no.~4, pp. 1558--2221, Apr. 2019.

\bibitem{MS-programmable}
K.~Chen~\emph{et al}, ``A reconfigurable active huygens' metalens,'' \emph{Adv.
  Mater.}, vol.~29, p. 1606422, 2017.

\bibitem{Wang_2021}
X.~Wang and C.~Caloz, ``Pseudo-random sequence ({PRS}) (space)time-modulated
  metasurfaces,'' \emph{engrXiv}, pp. 1--13, 2021.

\end{thebibliography}

\end{document}